\documentstyle[12pt]{article}
\newlength{\extraspace}
\newlength{\extraspaces}
\newcommand{\be}{\begin{equation}
\addtolength{\abovedisplayskip}{\extraspaces}
\addtolength{\belowdisplayskip}{\extraspaces}
\addtolength{\abovedisplayshortskip}{\extraspace}
\addtolength{\belowdisplayshortskip}{\extraspace}}
\newcommand{\ee}{\end{equation}}

\newcommand{\ba}{\begin{eqnarray}
\addtolength{\abovedisplayskip}{\extraspaces}
\addtolength{\belowdisplayskip}{\extraspaces}
\addtolength{\abovedisplayshortskip}{\extraspace}
\addtolength{\belowdisplayshortskip}{\extraspace}}
\newcommand{\ea}{\end{eqnarray}}

\newcommand{\nonu}{\nonumber \\[.5mm]}
\newcommand{\A}{&\!\!\!}

\newcommand{\newsection}[1]{
\vspace{7mm} \pagebreak[3] \addtocounter{section}{1}
\setcounter{subsection}{0} \setcounter{footnote}{0}
\begin{center}
{\large {\bf \thesection. #1}}
\end{center}
\nopagebreak
\medskip
\nopagebreak \hspace{3mm}}

\begin{document}

\begin{center}
{\bf Charged Dilaton,  Energy, Momentum and Angular-Momentum in
Teleparallel Theory Equivalent to General
Relativity}
\end{center}
\centerline{ Gamal G.L. Nashed}

\bigskip

\centerline{{\it Mathematics Department, Faculty of Science, Ain
Shams University, Cairo, Egypt }}

\bigskip
 \centerline{ e-mail:nasshed@asunet.shams.edu.eg}

\hspace{2cm}
\\
\\
\\
\\
\\

We apply the  energy-momentum tensor to calculate energy, momentum
and angular-momentum of two  different tetrad fields. This tensor
 is coordinate independent of the gravitational field
established in the Hamiltonian structure of the teleparallel
equivalent of general relativity (TEGR). The spacetime of these
tetrad fields is the charged dilaton. Our results show that the
 energy associated with one of these tetrad fields is consistent, while the
other one does not show this consistency. Therefore, we use the
regularized expression of the gravitational energy-momentum tensor
of the TEGR. We investigate the energy within the external event
horizon using  the definition of the gravitational
energy-momentum.
\begin{center}
\newsection{\bf Introduction}
\end{center}

Quantum mechanics and general relativity (GR) are two very
successful and well validated theories within their own domains.
The main problem is  to unify them into a single consistent
theory. One of the most promising models of unification is the
string theory. The string theory is classified into two classes
which are the closed and the open strings. The gravity is
described by the first class
 while the matter is described by the second one. In case of non-
 perturbative  string theory, there are extended objects known as
  D-branes. These objectives are surfaces where open strings must
  begin and finish which provide an alternative approach to the
  Kaluza-Klein \cite{KK}. In this last approach the matter penetrates
  the extra dimensions, leading to strong constraints from collider physics.

Nowadays there is a growing body of literature about the
gravitational field of string matter coupled to an
electromagnetically charged dilaton field. Black hole solutions in
dilaton gravity were first analyzed  by Gibbons and Maeda
\cite{GM1}. Garfinkle et al. \cite{GHS} have obtained a family of
solutions representing static, spherically symmetric charged black
holes. Kallosh and Peet \cite{KP} in the context of supersymmetric
theories investigated these solutions. When the dilaton acquires a
mass Gregory and Harvey \cite{GH1} modified the dilaton black
holes. A static spherically symmetric metric around a source
coupled to a massless dilaton with both electric and magnetic
charged is investigated by Agnese and Camera \cite{AC}

Among  various attempts to overcome the problems of quantization
and the existence of singular solution in Einstein's GR, gauge
theories of gravity are of special interest, as they based on the
concept of gauge symmetry which has been very successful in the
foundation of other fundamental interactions. The importance of
the Poincar$\acute{e}$ symmetry in particle physics leads one to
consider the Poincar$\acute{e}$ gauge theory (PGT) as a natural
framework for description of the gravitational phenomena
\cite{Kt}$\sim$\cite{HMM}.  Basic gravitational variables in PGT
are the tetrad field ${e^a}_\mu$ and the Lorentz connection
${A^{ab}}_\mu$. These variables  are associated to the translation
and Lorentz subgroups of the Poincar$\acute{e}$ group. The gauge
fields are coupled to the energy-momentum and spin of matter
fields, and their field strengths are geometrically identified
with the torsion and the curvature.

General geometric arena of PGT, the Riemann-Cartan space $U_4$,
may be a priori restricted by imposing certain conditions on the
curvature and the torsion. Thus, Einstein's GR is defined in
Riemann space $V_4$, which is obtained from $U_4$ by the
requirement of vanishing torsion. Another interesting limit of PGT
is the {\it teleparallel or Weitzenb$\ddot{o}$ck} geometry $T_4$.
The vanishing of the curvature means that parallel transport is
path independent. The teleparallel geometry is, in sense,
complementary to Riemannian: curvature vanishes, and torsion
remains to characterize the parallel transport. For the physical
interpretation of the teleparallel geometry  there is a
one-parameter family of teleparallel Lagrangians which is
empirically equivalent to GR \cite{HNV,HS9,Nj}. If the parameter
value $B=1/2$ the Lagrangian of the theory coincides, modulo a
four-divergence, with the Einstein-Hilbert Lagrangian, and defines
(TEGR).

The search for a consistent expression for the gravitating energy
and angular-momentum of a self-gravitating distribution of matter
is undoubtedly a long-standing problem in GR. It is believed that
the energy of the gravitational field is not localizable, i.e.,
defined in a finite region of the space. The gravitational field
does not possess the proper definition of an energy momentum
tensor. It is usually to define some energy-momentum and
angular-momentum \cite{BT,LL} which are pseudo-tensors and depend
on the second derivative of the metric tensor. These quantities
can be annulled by an adequate transformation of coordinate.
Bergmann \cite{BT}, Landau-Lifschitz \cite{LL} justify that
 the energy and angular momentum are  consistent with Einstein's principle of
equivalence. According to this principle ``any space-time region,
infinitesimal or not, is flat if and only if the
Riemann-Christoffel tensor vanishes in this region". In such a
flat space-time, energy of the gravitational field is null.
Therefore, it is only possible to define the energy of the
gravitational filed in whole space-time region and not only in a
small region. The Einstein's GR can also be reformulated in the
context of teleparallel geometry \cite{PP}$\sim$\cite{AGP1}. In
this geometry the dynamical field is corresponding to orthonormal
tetrad field ${e^a}_\mu$\footnote{space-time indices $\mu, \ \
\nu, \cdots$ and SO(3,1) indices a, b $\cdots$ run from 0 to 3.
Time and space indices are indicated to $\mu=0, i$, and $a=(0),
(i)$.} (a, $\mu$ are SO(3,1) and space-time indices,
respectively). The teleparallel geometry is a suitable framework
to address the notions of energy, momentum and angular-momentum of
any space-time that admits a $3+1$ foliation \cite{MR}. Therefore,
we consider the TEGR in this work.

In order to calculate the energy and angular momentum we use the
Hamiltonian that is formulated for {\it an arbitrary teleparallel
theories } using Schwinger's time gauge \cite{Sj}$\sim$
\cite{SM2}. In this formulation it is shown that the TEGR is the
{\it only viable consistent teleparallel theory of gravity}. Maluf
and Rocha \cite{MD} established a theory in which {\it Schwinger's
time gauge has not been incorporated in the geometry of absolute
parallelism}. In this formulation, the definition of the
gravitational angular-momentum arises by suitably interpreting the
integral form of the constraint equation $\Gamma^{ab}=0$. This
definition has been successfully applied to the gravitational
field of a thin, slowly rotating mass shell \cite{MUFR} and for
the three-dimensional BTZ black hole \cite{MPR}.

Definitions for the gravitational energy in the context of the
TEGR have already been proposed in the literatures \cite{NHC,Mj}.
An expressions for the gravitational energy arises from the
surface term of the total Hamiltonian is given \cite{Nj8, BV2001}.
These expressions are equivalent to the integral form of the total
divergences of the Hamiltonian density developed by Maluf et al.
\cite{MD}. These
 expressions yield the same value for the total energy of the
gravitational field. However, since these expressions contain the
lapse function in the integrand, non of them are suitable to the
calculation of the irreducible mass of the Kerr black hole. This
is  because the lapse function vanishes on the external event
horizon of the black hole \cite{Mj}. The energy expressions
\cite{Nj8, BV2001} neither to be applied to a finite surface
integration nor they yield the total energy of the space-time
\cite{Mj}. A good energy-momentum expression for gravitating
systems should satisfy a variety of requirements; to  give the
standard values of the total quantities for asymptotically flat
space, to reduce to the material energy-momentum in proper limit
and to be positive \cite{CN2,SNC}. No entirely expression has yet
been identified.  For more details of the topic of quasi-local
approach a review article is given \cite{SL}.

To calculate the energy and momentum, the definition of
energy-momentum, i.e., $P^a$, is given which is invariant under
global $SO(3,1)$ transformations. It has been argued elsewhere
\cite{Mj5} that $P^a$ makes sense to have a dependence on the
frame. The energy-momentum in classical theories of particles and
fields does not depend on the frame, and it has been asserted that
such dependence is a natural property of the gravitational
energy-momentum.  It is assumed that a set of tetrads fields is
adapted to an observer in the space-time determined by the metric
tensor $g_{\mu \nu}$.

We investigate the irreducible mass $M_{irr}$ of the dilaton black
hole. This $M_{irr}$ is the total mass of the black hole at the
final stage of Penrose's process of energy extraction, considering
that the maximum possible energy is extracted. The  $M_{irr}$  is
also related to the energy contained within the external event
horizon $E(r_+)$ of the black hole (the surface of the constant
radius $r=r_+$ defines the external event horizon). Every
expression for local or quasi-local gravitational energy must
necessary yield the value of $E(r_+)$ in close agreement with
$2M_{irr}$, since we know beforehand the value of $M_{irr}$  as a
function of the initial angular-momentum of the black hole
\cite{CD}. The evolution of $2M_{irr}$ is a crucial test for any
expression of the gravitational energy. $E(r_+)$ has been obtained
by means of different energy expressions \cite{BG}. The
gravitational energy used in this article is the only one that
yields a satisfactory value for $E(r_+)$ and that arises in the
framework of the Hamiltonian formulation of the gravitational
field.

The main aim of the present work is to reformulate the solution
given by  Garfinkle et al.  \cite{GHS} within the framework of
TEGR and then, compute  energy, momentum and  angular momentum
using the energy-momentum tensor. In \S 2 we briefly review the
TEGR theory for gravitational, electromagnetic and dilaton and
then we derive the equations of motion.  A summary of the
derivation of energy and angular-momentum is given in \S 3. In \S
4, we study the two tetrad fields and then calculate the energy
and angular-momentum. To calculate the energy associated with the
second tetrad field we use the regularized expression for the
gravitational energy-momentum in \S 5.  The final section is
devoted to discussion and conclusion.
\newsection{The  TEGR for gravitation, electromagnetic and dilaton}

In a space-time with absolute parallelism the parallel vector
fields ${e_a}^\mu$ define the nonsymmetric affine connection \be
{\Gamma^\lambda}_{\mu \nu} \stackrel{\rm def.}{=} {e_a}^\lambda
{e^a}_{\mu, \nu}, \ee where $e_{a \mu, \ \nu}=\partial_\nu e_{a
\mu}$\footnote{space-time indices $\mu, \ \ \nu, \cdots$ and
SO(3,1) indices a, b $\cdots$ run from 0 to 3. Time and space
indices are indicated to $\mu=0, i$, and $a=(0), (i)$.}. The
curvature tensor defined by ${\Gamma^\lambda}_{\mu \nu}$, given by
Eq. (1), is identically vanishing. The metric tensor $g_{\mu \nu}$
 is defined by
 \be g_{\mu \nu} \stackrel{\rm def.}{=}  \eta_{a b} {e^a}_\mu {e^b}_\nu, \ee
with $\eta_{a b}=(-1,+1,+1,+1)$ is the metric of Minkowski
space-time.

  The Lagrangian density for the gravitational field in the TEGR,
  in the presence of matter fields, is given by\footnote{Throughout this paper we use the
relativistic units$\;$ , $c=G=1$ and $\kappa=8\pi$.} \cite{Mj} \be
{\cal L}_G  =  e L_G =- \displaystyle {e \over 16\pi}  \left(
\displaystyle {T^{abc}T_{abc} \over 4}+\displaystyle
{T^{abc}T_{bac} \over 2}-T^aT_a
  \right)-L_m= - \displaystyle {e \over 16\pi} {\Sigma}^{abc}T_{abc}-L_m,\ee
where $e=det({e^a}_\mu)$. The tensor ${\Sigma}^{abc}$ is defined
by \be {\Sigma}^{abc} \stackrel {\rm def.}{=} \displaystyle{1
\over 4}\left(T^{abc}+T^{bac}-T^{cab}\right)+\displaystyle{1 \over
2}\left(\eta^{ac}T^b-\eta^{ab}T^c\right).\ee $T^{abc}$ and $T^a$
are the torsion tensor and the basic vector field  defined by \be
{T^a}_{\mu \nu} \stackrel {\rm def.}{=}
{e^a}_\lambda{T^\lambda}_{\mu
\nu}=\partial_\mu{e^a}_\nu-\partial_\nu{e^a}_\mu, \quad \quad
{T^a}_{b c} \stackrel {\rm def.}{=} {e_b}^\mu {e_c}^\nu {T^a}_{\mu
\nu}, \quad \quad T^a \stackrel {\rm def.}{=}{{T^b}_b}^a.\ee The
quadratic combination $\Sigma^{abc} T_{abc}$ is proportional to
the scalar curvature $R(e)$, except for a total divergence term
\cite{Mj}. $L_m$ represents the Lagrangian density for matter
fields.

The electromagnetic Lagrangian  density ${\it L_{e.m.}}$ is
\cite{KT}  \be {\cal L}_{e.m.}  =  e \ L_{e.m.} = e \ e^{-2\xi}
 g^{\mu \rho} g^{\nu \sigma} F_{\mu \nu} F_{\rho \sigma}, \ee with
 $F_{\mu \nu}$ being the
Maxwell field associated with a $U(1)$ subgroup of $E_8\times E_8$
and is defined by\footnote{Heaviside-Lorentz rationalized units
will be used throughout this paper} $F_{\mu \nu}\stackrel{\rm
def.}{=}
\partial_\mu A_\nu-\partial_\nu A_\mu$.

Finally the dilaton Lagrangian  density ${\it L_{D}}$ is
\cite{GHS} \be {\cal L}_D  =  e \ L_D = 2e \
\left(\bigtriangledown \xi \right)^2,  \ee with $\xi$ being the
dilaton.

The gravitational, electromagnetic and dilaton field equations for
the system described by ${\it L_G}+{\it L_{e.m.}}+L_{D}$ are the
following:

 \ba  \A \A e_{a \lambda}e_{b \mu}\partial_\nu\left(e{\Sigma}^{b \lambda \nu}\right)-e\left(
 {{\Sigma}^{b \nu}}_a T_{b \nu \mu}-\displaystyle{1 \over 4}e_{a \mu}
 T_{bcd}{\Sigma}^{bcd}\right)= \displaystyle{1 \over 2}{\kappa} eT_{a
 \mu},\nonu
\A \A \nabla_\mu\left(e^{-2\xi} F^{\mu \nu}\right)=0, \nonu
\A \A \nabla^2\xi+\frac{1}{2}e^{-2\xi}F^2=0,
 \ea
where \[ T_{\mu \nu}=2\left\{\nabla_\mu \xi \nabla_\nu
\xi-\frac{1}{2}g_{\mu \nu} g^{\rho \sigma}\nabla_\rho \xi
\nabla_\sigma \xi+e^{-2\xi}\left(g_{\nu \sigma}F_{\mu
\rho}F^{\sigma \rho}-\frac{1}{4}g_{\mu \nu}F^2 \right)\right\}.\]
It is possible to prove by explicit calculations that the left
hand side of the symmetric field equation (8) is exactly given by
\cite{Mj}
 \[\displaystyle{e \over 2} \left[R_{a
\mu}(e)-\displaystyle{1 \over 2}e_{a \mu}R(e) \right]. \] The
axial-vector part of the torsion tensor $a_\mu$ is defined by \be
a_\mu \stackrel{\rm def.}{=} {1 \over 6} \epsilon_{\mu \nu \rho
\sigma} T^{\nu \rho \sigma}={1 \over 3} \epsilon_{\mu \nu \rho
\sigma} \gamma^{\nu \rho \sigma}, \qquad where \qquad
\epsilon_{\mu \nu \rho \sigma} \stackrel{\rm def.}{=} \sqrt{-g}
\delta_{\mu \nu \rho \sigma}, \ee and $\delta_{\mu \nu \rho
\sigma}$ being completely antisymmetric and normalized as
$\delta_{0123}=-1$.

\newsection{Energy, momentum and angular-momentum}
In the context of Einstein's GR, rotational phenomena is certainly
not a completely understood issue. The prominent manifestation of
a purely relativistic rotation effect is the dragging of inertial
frames. If the angular-momentum of the gravitational field of
isolated system has a  meaningful notion, then it is reasonable to
expect the latter to be somehow related to the rotational motion
of the physical sources.

The angular-momentum of the gravitational field has been addressed
in the literature by means of different approaches. The oldest
approach is based on pseudotensors \cite{BT,LL}, out of which
angular-momentum superpotentials are constructed. An alternative
approach assumes the existence of certain Killing vector fields
that allow the construction of conserved integral quantities
\cite{Ka}. Finally, the gravitational angular-momentum can also be
considered in the context of Poincar$\acute{e}$ gauge theories of
gravity \cite{HS8}, either in the Lagrangian or in the Hamiltonian
formulation. In the latter case it is required that the generators
of spatial rotations at infinity have a well defined functional
derivatives. From this requirement a certain surface integral
arises, whose value is interpreted as the gravitational
angular-momentum.

The Hamiltonian formulation of TEGR is obtained by establishing
the phase space variables. The Lagrangian density does not contain
the time derivative of the tetrad component $e_{a0}$. Therefore,
this quantity will arise as a Lagrange multiplier \cite{Dp}. The
momentum canonically conjugated to $e_{ai}$ is given by
$\Pi^{ai}=\delta L/\delta \dot{e}_{ai}$. The Hamiltonian
formulation is obtained by rewriting the Lagrangian density in the
form $L=p \ \dot{q}-H$, in terms of $e_{ai}, \Pi^{ai}$ and the
Lagrange multipliers. The Legendre transformation can be
successfully carried out and the final form of the Hamiltonian
density has the form \cite{MR} \be H=e_{a0}C^a+\alpha_{ik}
\Gamma^{ik}+\beta_k\Gamma^k,\ee plus a surface term. Here
$\alpha_{ik}$ and $\beta_k$ are Lagrange multipliers that are
identified as \be \alpha_{ik}={1 \over 2} (T_{i0k}+T_{k0i}) \qquad
and \qquad \beta_k=T_{00k},\ee and $C^a$, $\Gamma^{ik}$ and
$\Gamma^k$ are first class constraints. The Poisson brackets
between any two field quantities $F$ and $G$ is given by \be \{
F,G \}=\int d^3x \left( \displaystyle{\delta F \over \delta
e_{ai}(x)} \displaystyle{\delta G \over \delta
\Pi^{ai}(x)}-\displaystyle{\delta F \over \delta
\Pi^{ai}(x)}\displaystyle{\delta G \over \delta e_{ai}(x)}
\right).\ee We recall that the Poisson brackets
$\left\{\Gamma^{ij}(x),\Gamma^{kl}(x)\right\}$ reproduce the
angular-momentum algebra \cite{Mj}.

 The constraint $C^a$ is
written as $C^a=-\partial_i \Pi^{ai}+h^a$, where $h^a$ is an
intricate expression of the field variables. The integral form of
the constraint equation $C^a=0$ motivates the definition of the
gravitational energy-momentum $P^a$ four-vector \cite{Mj} \be
P^a=-\int_V d^3 x
\partial_i \Pi^{ai},\ee where $V$ is an arbitrary volume of the
three-dimensional space. In the configuration space we have \ba
\Pi^{ai} \A =\A -4\kappa \sqrt{-g} \Sigma^{a0i} \quad with \quad
\partial_\nu(\sqrt{-g}\Sigma^{a \lambda \nu})=\displaystyle{1
\over 4 \kappa}\sqrt{-g}{e^a}_\mu (t^{\lambda \mu}+T^{\lambda
\mu}) \quad where \nonu
\A \A   t^{\lambda \mu}=\kappa \left(4\Sigma^{bc
\lambda}{T_{bc}}^\mu-g^{\lambda \mu} \Sigma^{bcd}T_{bcd}
\right).\ea

The emergence of total divergences in the form of scalar or vector
densities is possible in the framework of theories constructed out
of the torsion tensor. Metric theories of gravity do not share
this feature. By making $\lambda=0$ in Eq. (14) and identifying
$\Pi^{ai}$ in the left side of the latter, the integral form of
Eq. (13) is written as \be P^a=\int_V d^3 x \sqrt{-g}
{e^a}_\mu\left(t^{0 \mu}+T^{0 \mu} \right).\ee Eq. (15) suggests
that $P^a$ is now understood as the gravitational  energy-momentum
\cite{Mj}. The spatial component $P^{(i)}$ form a total
three-momentum, while temporal component $P^{(0)}$ is the total
energy \cite{LL}.

It is possible to rewrite the Hamiltonian density of Eq. (10) in
the equivalent form \cite{MUFR} \be H=e_{a0}C^a+\displaystyle{1
\over 2}\lambda_{ab}\Gamma^{ab}, \qquad with \qquad
\lambda_{ab}=-\lambda_{ba}, \ee are the Lagrangian multipliers
that are identified as $\lambda_{ik}=\alpha_{ik}$ and
$\lambda_{0k}=-\lambda_{k0}=\beta_k$.  The constraints
$\Gamma^{ab} = -\Gamma^{ba}$ \cite{MR} embodies both constraints
$\Gamma^{ik}$ and $\Gamma^k$ by means of the relation \be
\Gamma^{ik}={e_a}^i {e_b}^k \Gamma^{ab}, \qquad and \qquad
\Gamma^k \equiv \Gamma^{0k}={e_a}^0 {e_b}^k \Gamma^{ab}.\ee The
constraint $\Gamma^{ab}$ can be reads as \be
\Gamma^{ab}=M^{ab}+4\kappa\sqrt{-g}{e_{(0)}}^0
\left(\Sigma^{a(0)b}-\Sigma^{b(0)a}\right).\ee

In similarity to the definition of $P^a$, the integral form of the
constraint equation $\Gamma^{ab}=0$ motivates the new definition
of the space-time angular-momentum. The equation $\Gamma^{ab}=0$
implies \be M^{ab}=-4\kappa\sqrt{-g}{e_{c}}^0
\left(\Sigma^{acb}-\Sigma^{bca}\right),\ee Maluf et al. \cite{Mj,
MUFR} defined \be L^{ab} =\int_V d^3x {e_\mu}^a {e_\nu }^b M^{\mu
\nu }, \ee  as the 4-angular-momentum of the gravitational field
for an arbitrary volume V of the three-dimensional space. In
Einstein-Cartan type theories there also appear constraints that
satisfy the Poisson bracket as given by Eq. (12). However, such
constraints arise in the form $\Pi^{[ij]}=0$, and so a definition
similar to Eq. (20), i.e., interpreting the constraint equation as
an equation for the angular-momentum of the field, {\it is not
possible}. Definition (20) is three-dimensional integral. The
quantities $P^a$ and $L^{ab}$ are separately invariant under
general coordinate transformations of the three-dimensional space
and under time reparametrizations, which is an expected feature
since these definitions arise in the Hamiltonian formulation of
the theory. Moreover, these quantities transform covariantly under
global $SO(3,1)$ transformations \cite{MUFR}.

\newsection{Tetrad fields with spherical symmetry}

Now we will consider two simple configuration of  tetrad fields
and discuss their physical interpretation as reference frames. The
first one in quasi-orthogonal coordinate system  can be written as
\cite{Rh} \ba {{e_{(0)}}^0} \A= \A A, \quad {e_\alpha}^0 = C x^a,
\quad {e_{(0)}}^\alpha = D x^\alpha \nonu
{e_a}^\alpha \A= \A \delta_a^\alpha B + F x^a x^\alpha +
\epsilon_{a \alpha \beta} S x^\beta,
 \ea where {\it A}, {\it C},
{\it D}, {\it B}, {\it F}, and {\it S} are unknown functions of
${\it r}$. It can be shown that the unknown functions $D$ and $F$
can be eliminated by coordinate transformations \cite{HS7,SNH},
i.e., by making use of freedom to redefine $t$ and $r$, leaving
the tetrad field (21) having four unknown functions in the
quasi-orthogonal coordinates. Thus the tetrad field (21) without
the unknown functions $D$ and $F$ and also without the two unknown
functions $C$ and $S$ will be used in the following discussion for
the calculations of energy, momentum and angular-momentum but in
the spherical coordinate. Therefore, the tetrad field (21) can be
written in the spherical coordinates without the unknown functions
$D$, $F$, $C$ and $S$ as \cite{SNH} \be \left({{e_1}_a}^{ \mu}
\right) = \left( \matrix{ \frac{1}{A} &0 & 0 & 0 \vspace{3mm} \cr
0 & B \sin\theta \cos\phi & \frac{\cos\theta \cos\phi}{R(r)}
 & -\frac{\sin\phi} {R(r) \sin\theta} \vspace{3mm} \cr
0 & B \sin\theta \sin\phi & \frac{\cos\theta \sin\phi}{R(r)}
 & \frac{\cos\phi} {R(r) \sin\theta} \vspace{3mm} \cr
0 & B \cos\theta & -\frac{\sin\theta}{R(r)} & 0 \cr } \right). \ee

The other configuration of tetrad field that has a simple
interpretation as a reference frame can has the form \be
\left({{e_2}_a}^{ \mu} \right) = \left( \matrix{ \frac{1}{A} &0 &
0 & 0 \vspace{3mm} \cr 0 & B & 0
 & 0 \vspace{3mm} \cr
0 & 0 & \frac{1}{R(r)}
 & 0 \vspace{3mm} \cr
0 & 0 & 0&\frac{1}{R(r) \sin\theta}  \cr } \right). \ee The two
tetrads (22) and (23) are related by a local Lorentz
transformation which keeps spherical symmetry, i.e., the tetrad
(22) can be written in terms of the tetrad (23) using the
following local Lorentz transformation \be \left({{e_1}_a}^{ \mu}
\right) = {\Lambda_\nu}^\mu \left({{e_2}_a}^{ \nu} \right), \
where \ {\Lambda_\nu}^\mu=\left( \matrix{ 1&0 & 0 & 0 \vspace{3mm}
\cr 0 & \sin\theta \cos\phi & \cos\theta \cos\phi
 & -\sin\phi \vspace{3mm} \cr
0 &  \sin\theta \sin\phi & \cos\theta \sin\phi
 & \cos\phi \vspace{3mm} \cr
0 &  \cos\theta & -\sin\theta & 0 \cr } \right). \ee

The space-time associated with the two tetrad fields (22) and (23)
is the same and has the form \be ds^2=-A^2 dt^2+\frac{1}{B^2}
dr^2+R(r)^2(d\theta^2+\sin^2\theta d\phi^2).\ee

 Now we are going to calculate the energy, momentum and angular-momentum
 associated with the two tetrad fields (22) and (23). For asymptotically flat
space-times $P^0$ yields the ADM energy \cite{ADM}. In the context
of tetrad theories of gravity, asymptotically flat space-times may
be characterized by the asymptotic boundary condition \be e_{a
\mu} \cong \eta_{a \mu} + \displaystyle{1 \over 2} h_{a
\mu}(1/r),\ee and by the condition $\partial_\mu
{e^a}_\mu=O(1/r^2)$ in the asymptotic limit $r \rightarrow
\infty$. An important property of tetrad fields that satisfy Eq.
(26) is that in the flat space-time limit one has
${e^a}_\mu(t,x,y,z)={\delta^a}_\mu$, and therefore the torsion
tensor ${T^a}_{\mu \nu}=0$.

 Now we are going to apply Eq. (13) to the tetrad field (22) to calculate
 the energy content. We perform the calculations in the spherical coordinate.
  Eqs. (22) and (23)
 assumed that the reference space is determined by a set of tetrad
 fields ${e^a}_\mu$ for the flat space-time such that the
 condition ${T^a}_{\mu \nu}=0$ is satisfied. Using Eq. (5) in Eq.
 (22), the non-vanishing components of the torsion tensor  are
given by \ba {T^{(0)}}_{01} \A=\A \frac{A'}{A}, \qquad \qquad
{T^{(2)}}_{12}=\frac{(1-R'(r)B)}{R(r)B}={T^{(3)}}_{13},\ea and the
non-vanishing component of the tensor $T^{(a)}$ defined by Eq. (5)
is given by \be T^{(1)}=\frac{B(r)
\left\{2A(r)-2R'(r)A(r)B(r)-A'(r)B(r) R(r) \right\}}{R(r)A}. \ee
The axial vector associated with Eq. (22) is vanishing identically
due to the fact that the tetrad field of Eq. (22) has a spherical
symmetry \cite{Rh}.

Now we are going to apply Eq. (13) to the tetrad field (22) using
Eqs. (27) and (28) to calculate
 the energy content. We perform the calculations in the spherical
 coordinate. The only required component of ${\Sigma}^{\mu \nu
 \lambda}$ is
 \be
{\Sigma}^{(0) 0 1}=-\frac{R(r) \sin\theta \{1-R'(r)B\}}{4\pi }.\ee
Using Eq. (29) in (13) we obtain \be P^{(0)}=E=-\oint_{S
\rightarrow \infty}
 dS_k \Pi^{(0) k}=-\displaystyle {1  \over 4 \pi} \oint_{S \rightarrow \infty}
 dS_k  e
{\Sigma}^{(0) 0 k}= R(r){\{1-R'(r)B\}}.\ee Now let us apply
expression (13) to the evaluation of the irreducible mass by
fixing $V$ to be the volume within the $r=r_+$ surface where $r_+$
is the external horizon, i.e., $B=0$. Therefore, \be
P^{(0)}=E=-\int_{S_i} dS_i \Pi^{(0)i}=-\int_S d\theta d\phi
\Pi^{(0)1}(r,\theta,\phi),\ee where the surface $S$ is determined
by the condition $r=r_+$. The expression of $\Pi^{(0)1}$ will be
obtained by considering Eq. (14) using Eq. (4) and Eq. (5). The
expression of $\Pi^{(0)1}(r,\theta,\phi)$ for the tetrad (22)
 reads \be \Pi^{(0)1}(r,\theta,\phi)=\frac{\sin\theta R(r_+)\{1-R'(r_+)B(r_+)\}}{4\pi
}, \ee integrate Eq. (32) on the surface of constant radius
$r=r_+$ where $r_+$ is the external horizon of the  black hole. On
this surface the second term of Eq. (32) vanishes, i.e.,
$B(r_+)=0$. Therefore, on the surface $r=r_+$ we get \be
P^{(0)}=E=R(r_+).\ee Eq. (33) consistent  with the results
obtained before when $R(r_+)=r_+$ \cite{Mj,SL}, otherwise we
obtained a different result. It is shown \cite{GHS}  that the
unknown functions in the  metric given by Eq. (25) may have the
value \be A  = \frac{1}{B}=\sqrt{1-\frac{2M}{r}}, \qquad \qquad
R(r)=r\sqrt{1-\frac{Q^2 e^{-2\xi_0}}{rM}}.\ee  For the value of
the unknown functions given by Eq. (34) to satisfy the field
equation (8) the dilaton, the vector potential, the Maxwell field
and the energy momentum tensor must have the form \ba \xi \A=\A
\xi_0-\frac{1}{2}ln \left(1-\frac{Q^2 e^{-2\xi_0}}{rM}\right),
\qquad A_3=Q\cos\theta, \qquad F_{23}=Q\sin\theta, \nonu
 {T^0}_0 \A=\A \frac{Q^2\left(4rM^2e^{2\xi_0}
-6MQ^2+rQ^2\right)}{4r^3\left( rM e^{2\xi_0}-Q^2 \right)^2},
\qquad {T^1}_1 = \frac{Q^2\left(4rM^2e^{2\xi_0}
-2MQ^2-rQ^2\right)}{4r^3\left( rM e^{2\xi_0}-Q^2 \right)^2} \nonu
{T^2}_2 \A=\A {T^3}_3 = -\frac{Q^2\left(4r^2M^3e^{4\xi_0}
-6rM^2Q^2e^{2\xi_0}+2MQ^4-r^2MQ^2e^{2\xi_0}+rQ^4\right)}{4r^3\left(
rM e^{2\xi_0}-Q^2 \right)^3},\ea where $M, Q, \xi, \xi_0$ are the
mass, charge, dilaton and the asymptotic value of the dilaton
respectively.  As is clear  from Eq. (33) that if the
 charge $Q=0$ then $R(r)=r$ and  the irreducible mass
 will coincides with that obtained  before \cite{Mj,SL}.
 Equation (33) tell us that the energy associated with the solution given by
 Eq. (34) on  the surface $R(r_+)$ is different from what is well
 known \cite{Mj,SL}.  To over
come such problem let us use a coordinate transformation which
makes $R(r)$ that appear in Eq. (34) to be $r$.

Now we are going to redefine the radial coordinate to be \be
r=\sqrt{R^2+\frac{Q^4e^{-4\xi_0}}{4M^2}}+\frac{Q^2e^{-2\xi_0}}{2M}.\ee
Using the coordinate transformation (36) in Eq. (22) we get \be
\left({{e_1}_a}^{ \mu} \right) = \left( \matrix{
\displaystyle{\frac{1}{\sqrt{1-\frac{4M^2e^{2\xi}}{\lambda(R)}}}}
&0 & 0 & 0 \vspace{3mm} \cr 0 & \displaystyle{\frac{\sigma(R)
\sqrt{1-\frac{4M^2e^{2\xi}}{\lambda(R)}} \sin\theta \cos\phi
}{2MR}} & \frac{\cos\theta \cos\phi}{R}
 & -\frac{\sin\phi} {R \sin\theta} \vspace{3mm} \cr
0 &
\displaystyle{\frac{\sigma(R)\sqrt{1-\frac{4M^2e^{2\xi}}{\lambda(R)}}\sin\theta
\sin\phi}{2MR}} & \frac{\cos\theta \sin\phi}{R}
 & \frac{\cos\phi} {R \sin\theta} \vspace{3mm} \cr
0 & \displaystyle{\frac{\sigma(R)
\sqrt{1-\frac{4M^2e^{2\xi}}{\lambda(R)}}\cos\theta}{2MR}} &
-\frac{\sin\theta}{R} & 0 \cr } \right), \ee where \be \lambda(R)
\stackrel{\rm def.}{=}
\sqrt{4R^2M^2+Q^4e^{-4\xi_0}}e^{2\xi_0}+Q^2, \qquad \sigma(R)
\stackrel{\rm def.}{=} \sqrt{4R^2M^2+Q^4e^{-4\xi_0}}, \ee and the
associated spacetime of Eq. (37) is given by
 \be ds^2=-\left(1-\frac{4M^2e^{2\xi}}{\lambda(R)}\right)dt^2+\frac{4M^2R^2}
 {\sigma^2(R)\left({1-\frac{4M^2e^{2\xi}}{\lambda(R)}}\right)}
dR^2+R^2(d\theta^2+\sin^2\theta d\phi^2).\ee When the dilaton
$\xi$ and the charge $Q$ are vanishing  then, Eq. (37) will be
identical with the tetrad field that reproduce Schwarzschild
spacetime \cite{NS}. When the dilation solution is vanishing then
Eq. (37) will behave asymptotically like

\be \left({{e_1}_a}^{ \mu} \right) \cong \left( \matrix{
\frac{2R^2+2MR+3M^2-Q^2}{2R^2} &0 & 0 & 0 \vspace{3mm} \cr 0 &
\frac{ \sin\theta \cos\phi(2R^2-2MR-M^2+Q^2)}{2R^2}
 & \frac{\cos\theta \cos\phi}{R}
 & -\frac{\sin\phi} {R \sin\theta} \vspace{3mm} \cr
0 & \frac{ \sin\theta \sin\phi(2R^2-2MR-M^2+Q^2)}{2R^2}&
\frac{\cos\theta \sin\phi}{R}
 & \frac{\cos\phi} {R \sin\theta} \vspace{3mm} \cr
0 & \frac{ \cos\theta(2R^2-2MR-M^2+Q^2)}{2R^2} &
-\frac{\sin\theta}{R} & 0 \cr } \right), \ee and the associated
metric of Eq. (40) has the form \be ds^2 \cong
-\left(1-\frac{2MR-Q^2}{R^2}\right)dt^2+\left(1+\frac{2MR-Q^2}{R^2}\right)
dR^2+R^2(d\theta^2+\sin^2\theta d\phi^2),\ee which is the
asymptotic form of Reissner--Nordstr$\ddot{o}$m metric \cite{NS}.

Repeat the calculations of energy using the tetrad field given by
Eq. (37). Using Eq. (5) in Eq.
 (37), the non-vanishing components of the torsion tensor  are
given by \ba {T^{(0)}}_{01} \A=\A \frac{2\lambda'(R)
M^2e^{2\xi_0}}{\lambda(R)[\lambda(R)-4M^2e^{\xi_0}]}, \nonu
{T^{(2)}}_{12}\A=\A
\frac{\sigma(R)\sqrt{\lambda(R)-4M^2e^{2\xi_0}}-2MR\sqrt{\lambda(R)}}
{R \sigma(R) \sqrt{\lambda(R)-4M^2e^{2\xi_0}}}={T^{(3)}}_{13},\ea
and the non-vanishing component of the tensor $T^{(a)}$ is given
by \be
T^{(1)}=\frac{-\sigma(R)\left\{\sigma(R)M^2e^{2\xi_0}(4\lambda(R)-R\lambda'(R))+
2RM\sqrt{\lambda^4(R)-4\lambda^3(R)M^2e^{2\xi_0}}-\lambda^2(R)\sigma(R)\right\}}
{2R^3M^2\lambda^2(R)}. \ee

Using Eqs. (42) and (43) to calculate
 the energy content.  The only required component of ${\Sigma}^{\mu \nu
 \lambda}$ is
 \be
{\Sigma}^{(0) 0 1}=-\frac{ \sin\theta
\{2MR\sqrt{\lambda(R)}-\sigma(R)\sqrt{\lambda(R)-4M^2e^{2\xi_0}}\}}{8M\pi
\sqrt{\lambda(R)}}.\ee Substituting Eq. (44) in (13) we
obtain\newpage \ba P^{(0)} \A=\A E=-\oint_{S \rightarrow \infty}
 dS_k \Pi^{(0) k}=-\displaystyle {1  \over 4 \pi} \oint_{S \rightarrow \infty}
 dS_k  e{\Sigma}^{(0) 0
k}=\frac{\{2MR\sqrt{\lambda(R)}-\sigma(R)\sqrt{\lambda(R)-4M^2e^{2\xi_0}}\}}{2M
\sqrt{\lambda(R)}}\nonu
\A \A =
R-\frac{e^{-2\xi_0}\left(\sqrt{4R^2M^2e^{4\xi_0}+Q^4}\sqrt{\sqrt{4M^2R^2e^{4\xi_0}+Q^4}+Q^2-
4M^2e^{2\xi_0}}\right)}{2M
\sqrt{\sqrt{4M^2R^2e^{4\xi_0}+Q^4}+Q^2}},\nonu
\A \A \cong
M-\frac{4Q^2M^2e^{-2\xi_0}-4M^4+Q^4e^{-4\xi_0}}{8M^2R}+O\left(\frac{1}{R^2}\right),\ea
where we have used the definitions of $\lambda(R)$ and $\sigma(R)$
given
 by Eq. (38). For large $R$, i.e., $limit_{R \rightarrow \infty}$,  Eq. (45) will
 give the ADM \cite{MTW}.  If the asymptotic dilaton $\xi_0$ is
vanishing then the asymptotic form of the energy can be obtain
from Eq. (45) to have the value \be E \cong
M-\frac{4Q^2M^2-4M^4+Q^4}{8M^2R},\ee which is the energy of
Reissner-Nordstr$\ddot{o}$m space-time when $Q^4=0$ and $M^2=0$
\cite{NS}.

 Apply expression (13)
to the evaluation of the irreducible mass by fixing $V$ to be the
volume within the $R=R_+$ surface where $R_+$ is the external
horizon, i.e., $\left(1-\frac{4M^2e^{2\xi}}{\lambda(R_+)}
\right)\sigma^2(R_+)$=0. Therefore, \be P^{(0)}=E=-\int_{S_i} dS_i
\Pi^{(0)i}(R,\theta,\phi)=-\int_S d\theta d\phi
\Pi^{(0)1}(R,\theta,\phi),\ee where the surface $S$ is determined
by the condition $R=R_+$. The expression of $\Pi^{(0)1}$ will be
obtained by considering Eq. (14) using Eqs. (4) and (5). The
expression of $\Pi^{(0)1}(R,\theta,\phi)$ for the tetrad (37)
 reads \be \Pi^{(0)1}(R,\theta,\phi)=\frac{\{2MR_+\sqrt{\lambda(R_+)}-\sigma(R_+)
 \sqrt{\lambda(R_+)
 -4M^2e^{2\xi_0}}\}}{8M\pi
\sqrt{\lambda(R_+)} }, \ee where $R_+$ is the external horizon of
the black hole. On this surface the second term of Eq. (48)
vanishes, i.e., $\sqrt{\lambda(R_+) -4M^2e^{2\xi_0}}\sigma(R_+)$.
Therefore, on the surface $R=R_+$ integration of Eq. (48) will
give \be P^{(0)}=E=R_+,\ee which is a satisfactory  result that is
obtained before \cite{Mj,SL}.

Using Eq. (14) in Eq. (37)  to calculate the
 momentum and angular-momentum associated with the first tetrad
field given by Eq. (37). In this case
  we get \be \Pi^{(1) 1}(R,\theta,\phi)=0.\ee Substitute Eq. (50)
 in Eq. (13) we
get \be P^{(1)} =\int_VdV
\partial_{1}(\Pi^{(1)1}(R,\theta,\phi))=
\int_S dS_1 \Pi^{(1)1}(R,\theta,\phi) = 0 .\ee  By the same method
we obtain \be \Pi^{(2) 1}(R,\theta,\phi) =  0,   \qquad P^{(2)}= 0
,\qquad \Pi^{(3) 1}(R,\theta,\phi) = 0, \qquad P^{(3)}= 0 . \ee
The results of Eqs. (51) and (52) are expected results since the
space-time given by Eq. (37) is a spherically symmetric  static
space-time. Therefore, the spatial momentum associated with any
static solution is identically vanishing \cite{MTW}.

We use Eqs. (19) and Eq. (4) in Eq. (20) to calculate the
components of the angular-momentum. Finally we get \ba
M^{(0)(1)}(R,\theta,\phi) \A=\A \frac{-R\sin^2\theta \cos\phi{\
\{2MR\sqrt{\lambda^2(R)-4M^2\lambda(R)e^{2\xi_0}}+\sigma(R)\lambda(R)-4\sigma(R)M^2e^{2\xi_0}\}}}{4\pi
\sigma(R)\lambda(R)}, \nonu
 M^{(0)(2)} (R,\theta,\phi)\A=\A \frac{-R\sin^2\theta \sin\phi{\
\{2MR\sqrt{\lambda^2(R)-4M^2\lambda(R)e^{2\xi_0}}+\sigma(R)\lambda(R)-4\sigma(R)M^2e^{2\xi_0}\}}}{4\pi
\sigma(R)\lambda(R)},\nonu
 M^{(0)(3)} (R,\theta,\phi) \A =\A \frac{-R\sin\theta \cos\theta{\
\{2MR\sqrt{\lambda^2(R)-4M^2\lambda(R)e^{2\xi_0}}+\sigma(R)\lambda(R)-4\sigma(R)M^2e^{2\xi_0}\}}}{4\pi
\sigma(R)\lambda(R)},\nonu
 M^{(1)(2)} (R,\theta,\phi) \A =\A M^{(1)(3)}(R,\theta,\phi)= M^{(2)(3)}(R,\theta,\phi)=0.\ea
 Using Eq. (53) in Eq. (20) we get
 \ba L^{(0)(1)} \A =\A {\int_0^\pi}{\int_0^{2\pi}}{\int_{0}^\infty}
 d\theta d\phi dR M^{(0)(1)}(R,\theta,\phi) = 0 ,\nonu
 \A \A by \ \ \ the \ \ \  same \ \ \  method \ \ \  we \ \ \  can \ \ \
 get \nonu
 L^{(0)(2)} \A=\A L^{(0)(3)} =
 L^{(1)(2)}=L^{(1)(3)} = L^{(2)(3)}= 0.\ea It is of interest to note that
the vanishing of $L^{(0)(1)}$, $L^{(0)(2)}$ is due to the
appearance of terms like  $\sin\phi$ and $\cos\phi$ while the
vanishing of $L^{(0)(3)}$ is due to the
 appearance of term like $\sin\theta \cos \theta$.

To repeat the same computation for the tetrad (23) it is
sufficient to use the rules of transformation of the conserved
quantities \cite{OBRU}. Therefore, the  required component of
${\Sigma}^{\mu \nu \lambda}$ needed to calculate the energy  of
the tetrad field (23) has the form
 \be
{\Sigma}^{(0) 0 1}=-\frac{B(r)R(r)R'(r) \sin\theta}{4\pi}.\ee
Substituting (55) in (13) we obtain \ba P^{(0)}\A=\A E=-\oint_{S
\rightarrow \infty} dS_k \Pi^{(0) k}(r,\theta,\phi)=-\displaystyle
{1 \over 4 \pi} \oint_{S \rightarrow \infty}
 dS_k  e{\Sigma}^{(0) 0 k}=-B(r)R(r)R'(r)\nonu
\A=\A -\frac{(2rM-Q^2e^{-2\xi_0})\sqrt{1-\frac{2M}{r}}}{2M}.\ea
When the asymptotic dilaton $\xi_0=0$ and the charge $Q=0$ then
the asymptotic form of the above form of energy is given by \be E
\cong M-r, \ee which is different from the ADM form \cite{MTW}.
This is  due to the fact that the components of the torsion
 when $M=0$, $Q=0$ and $\xi_0=0$ do not
vanishing identically contradict the flatness condition given by
Eq. (26). Therefore, in this case we are going to use the
regularized expression for the gravitational energy-momentum
\cite{Mj}.
\newsection{Regularized expression for the gravitational energy-momentum
  and localization of energy}

An important property of the tetrad fields that satisfy the
condition of Eq. (26) is that in the flat space-time limit
${e^a}_\mu(t,x,y,z)={\delta^a}_\mu$, and therefore the torsion
${T^\lambda}_{\mu \nu}=0$.  Hence for the flat space-time it is
normally to consider a set of tetrad fields such that
${T^\lambda}_{\mu \nu}=0$ {\it in any coordinate system}. However,
in general an arbitrary set of tetrad fields that yields the
metric tensor for the asymptotically flat space-time does not
satisfy the asymptotic condition given by (26). Moreover for such
tetrad fields the torsion ${T^\lambda}_{\mu \nu} \neq 0$ for the
flat space-time \cite{MVR}. It might be argued, therefore, that
the expression for the gravitational energy-momentum (13) is
restricted to particular class of tetrad fields, namely, to the
class of frames such that ${T^\lambda}_{\mu \nu}=0$ if ${E_a}^\mu$
represents the flat space-time tetrad field \cite{MVR}. To explain
this, let us calculate the flat space-time of the tetrad field of
Eq. (23) using (34) which is given by \be \left({{E_2}_a}^\mu
\right) =\left(\matrix {1&0 &0 &0 \vspace{3mm} \cr 0&1 &0& 0
\vspace{3mm} \cr 0& 0&\frac{1}{r}&0 \vspace{3mm} \cr
0&0&0&\frac{1}{r\sin\theta} \cr } \right). \ee Expression (58)
yields the following non-vanishing torsion components: \be
{T^{(2)}}_{12}=-\frac{1}{r}={T^{(3)}}_{13}, \qquad \qquad
{T^{(3)}}_{23}=- \cot \theta.\ee The tetrad field (58) when
written in the Cartesian coordinate will have the form \be
\left({{E_2}_a}^\mu(t,x,y,z) \right) =\left(\matrix {1&0 &0 &0
\vspace{3mm} \cr 0& \frac{x}{r} & \frac{y}{r} & \frac{z}{r}
\vspace{3mm} \cr 0 & \frac{xz}{r\sqrt{x^2+y^2}} &
\frac{yz}{r\sqrt{x^2+y^2}}& -\frac{\sqrt{x^2+y^2}}{r} \vspace{3mm}
\cr 0&-\frac{y}{\sqrt{x^2+y^2}}&\frac{x}{\sqrt{x^2+y^2}}& 0 \cr }
\right). \ee In view of the geometric structure of Eq. (60), we
see that, Eq. (23) does not display the asymptotic behavior
required by Eq. (26). Moreover, in general the tetrad field (60)
is adapted to accelerated observers \cite{MR,Mj,MVR}. To explain
this, let us consider a boost in the x-direction of Eq. (60).  We
find \be \left({{E_2}_a}^\mu(t,x,y,z) \right) =\left(\matrix
{\gamma&v \gamma &0 &0 \vspace{3mm} \cr \frac{v \gamma x}{r} &
\frac{\gamma x}{r} &\frac{y}{r} &\frac{z}{r} \vspace{3mm} \cr
\frac{v \gamma xz}{r\sqrt{x^2+y^2}}&\frac{\gamma
xz}{r\sqrt{x^2+y^2}} &\frac{
yz}{r\sqrt{x^2+y^2}}&\frac{-\sqrt{x^2+y^2}}{r} \vspace{3mm} \cr
\frac{- v \gamma y}{\sqrt{x^2+y^2}}&\frac{-\gamma
y}{\sqrt{x^2+y^2}}&\frac{x}{\sqrt{x^2+y^2}}& 0 \cr } \right), \ee
where $v$ is the speed of the observer and
$\gamma=\frac{1}{\sqrt{1-v^2}}$. For a static object in a
space-time whose four-velocity is given by $u^\mu=(1,0,0,0)$ we
may compute its frame components $u^a={e^a}_\mu u^\mu=(\gamma,
\frac{v \gamma x}{r},\frac{v \gamma xz}{r\sqrt{x^2+y^2}},\frac{- v
\gamma y}{\sqrt{x^2+y^2}})$. It can be shown that along an
observer's trajectory whose velocity is determined by $u^a$  the
quantities \be {\phi_{(j)}}^{(k)}=u^i\left({{E_2}^{(k)}}_m
\partial_i {{E_2}_{(j)}}^m\right),\ee constructed out from Eq. (61) are
non vanishing. This fact indicates that along the observer's path
the spatial axis ${{E_2}_{(a)}}^\mu$ rotate \cite{MR,MVR}. In
spite of the above problems discussed for the tetrad field of Eq.
(23) it yields a satisfactory value for the total gravitational
energy-momentum, as we will discussed.

 In Eq. (13) it is implicitly assumed that the reference space is determined
 by a set of tetrad fields ${e^a}_\mu$ for flat space-time such
 that the condition ${T^a}_{\mu \nu}=0$ is satisfied. However, in
 general there exist flat space-time tetrad fields for which ${T^a}_{\mu \nu} \neq
 0$. In this case Eq. (13) may be generalized \cite{MR,MVR} by
 adding a suitable reference space subtraction term, exactly like
 in the Brown-York formalism \cite{BHS,YB}.

 We will denote ${T^a}_{\mu \nu}(E)=\partial_\mu {E^a}_\nu-\partial_\nu
 {E^a}_\mu$ and $\Pi^{a j}(E)$ as the expression of $\Pi^{a j}$
 constructed out of the flat tetrad ${E^a}_\mu$. {\it The
 regularized form of the gravitational energy-momentum $P^a$ is
 defined by} \cite{MR,MVR}
 \be P^a=-\int_{V} d^3x \partial_k \left[ \Pi^{a k}(e)-\Pi^{a k}(E)
 \right].\ee This condition guarantees that the energy-momentum of
 the flat space-time always vanishes. The reference space-time is
 determined by tetrad fields ${E^a}_\mu$, obtained from
 ${e^a}_\mu$ by requiring the vanishing of the physical parameters
 like mass, angular-momentum, etc. Assuming that the space-time is
 asymptotically flat then Eq. (63) can have the form \cite{MR,MVR}
\be P^a=-\oint_{S\rightarrow \infty} dS_k \left[ \Pi^{a
k}(e)-\Pi^{a k}(E) \right],\ee where the surface $S$ is
established at spacelike infinity. Eq. (64) transforms as a vector
under the global SO(3,1) group \cite{Mj}.

We may likewise establish the regularized expression for the
gravitational 4-angular momentum. It reads \be L^{ab}={\int_V}
d^3x \left[M^{ab}(e)-M^{ab}(E)\right].\ee

Now we are in a position to proof that the tetrad field (23)
yields a satisfactory value for the total gravitational
energy-momentum.  We will integrate Eq. (64) over a surface of
constant radius $x^1=r$ and require $r\rightarrow \infty$.
Therefore, the index $k$ in (64) takes the value $k=1$. We need to
calculate the quantity
\[\Sigma^{(0) 01}={e^{(0)}}_0\Sigma^{0 01}=\displaystyle{1 \over
2}{e^{(0)}}_0(T^{001}-g^{00}T^{1}).\] Evaluate  the above equation
we find \be \Pi^{(0)1}(e)=\displaystyle{-1 \over 4\pi}e\Sigma^{(0)
01}=-\frac{\sin\theta(2rM-Q^2e^{-2\xi_0})\sqrt{1-\frac{2M}{r}}}{8\pi
M},\ee and the expression of $\Pi^{(0)1}(E)$ is obtained by just
making $M=0$, $Q=0$ and $\xi_0=0$ in Eq. (66), it is given by \be
\Pi^{(0)1}(E)=\displaystyle{-1 \over 4\pi}r\sin \theta.\ee Thus
the gravitational energy of the tetrad field of Eq. (23) is given
by \ba P^{(0)}\A=\A \int d\theta d\phi \displaystyle{1 \over 4\pi}
\sin \theta
\left(r-\frac{(2rM-Q^2e^{-2\xi_0})\sqrt{1-\frac{2M}{r}}}{2
M}\right),\nonu
\A \A r-\frac{(2rM-Q^2e^{-2\xi_0})\sqrt{1-\frac{2M}{r}}}{2M} \cong
M+ \frac{Q^2 e^{-2\phi_0}}{2M} +O\left(\frac{1}{r}\right).\ea
which is exactly the ADM when $Q^2=0$ up
to $O(1/r)$. Eq. (68) tells us that when \\
$(2rM-Q^2e^{-2\xi_0})=0$ the form of the energy given by Eq. (68)
will effect and this is one of the defect of the solution given by
Eq. (34) \cite{GHS}. Therefore, we use the coordinate
transformation given by Eq. (36). The tetrad (23) after using
transformation  (36) will have the form \be \left({{e_2}_a}^{ \mu}
\right) = \left( \matrix{
\displaystyle{\frac{1}{\sqrt{1-\frac{4M^2e^{2\xi}}{\lambda(R)}}}}
&0 & 0 & 0 \vspace{3mm} \cr 0 & \displaystyle{\frac{\sigma(R)
\sqrt{1-\frac{4M^2e^{2\xi}}{\lambda(R)}}}{2MR}} & 0
 &0 \vspace{3mm} \cr
0 & 0 & \frac{1}{R}
 &0 \vspace{3mm} \cr
0 & 0& 0&\frac{1}{R\sin\theta}& \cr } \right). \ee Repeat the
calculations done above the non-vanishing components of the
torsion tensor and the vector field $T^{(a)}$ of the tetrad field
given by Eq. (69) have the form \ba {T^{(0)}}_{01} \A =\A
-\frac{2M^2\lambda'(R)e^{2\xi_0}}{\lambda(R)[\lambda(R)-4M^2e^{2\xi_0}]},
\quad  \quad {T^{(2)}}_{12}={T^{(3)}}_{13}=-\frac{1}{R},\quad
\quad {T^{(3)}}_{23}=- \cot \theta, \nonu
T^{(1)} \A=\A
-\frac{\sigma(R)\{\lambda'(R)-4\lambda(R)M^2e^{2\xi_0}-RM^2\lambda'(R)e^{2\xi_0}\}}
{R^3M^2\lambda^2(R)}, \qquad T^{(2)}=-\frac{\cot \theta}{R^2}.\ea

The only required component of ${\Sigma}^{\mu \nu
 \lambda}$ needed to calculate the energy  using the regularized expression given by Eq. (64)
 is \be {\Sigma}^{(0) 0
1}(e)=\frac{\left(\frac{\sigma(R)\sqrt{\lambda(R)-4M^2e^{2\xi_0}}}{2M\sqrt{\lambda(R)}}\right)
\sin\theta}{4\pi}, \qquad {\Sigma}^{(0) 0 1}(E)=\frac{R
\sin\theta}{4\pi}.\ee Substituting (71) in (64) we obtain \ba
P^{(0)}\A=\A E=-\oint_{S \rightarrow \infty}
 dS_k \Pi^{(0) k}(R,\theta,\phi)=-\displaystyle {1  \over 4 \pi} \oint_{S \rightarrow \infty}
 dS_k  e{\Sigma}^{(0) 0 k},\nonu
\A=\A
R-\frac{\sigma(R)\sqrt{\lambda(R)-4M^2e^{2\xi_0}}}{2M\sqrt{\lambda(R)}}\nonu
\A=\A R-
\frac{\sqrt{4M^2R^2+Q^4e^{-4\xi_0}}\sqrt{\sqrt{4M^2R^2+Q^4e^{-4\xi_0}}e^{2\xi_0}+Q^2-
4M^2e^{2\xi_0}}}{2M\sqrt{\sqrt{4M^2R^2+Q^4e^{-4\xi_0}}e^{2\xi_0}+Q^2}}\nonu
\A\A \cong M+O\left(\frac{1}{R}\right), \ which \ is \ the \ ADM \
up \ to \ O\left(\frac{1}{R}\right), \nonu
\A \A \cong M-\frac{4Q^2
M^2e^{-2\xi_0}+4M^4-Q^4e^{-4\xi_0}}{8M^2R}+O\left(\frac{1}{R^2}\right),
\ which \ is \ the \ energy \ of \ Reissner-Nordstr\ddot{o}m \nonu
\A \A space-time \ when \ the \ asymptotic \ dilaton \  \xi_0=0, \
\ Q^4=0 \ and \ M^2=0 \ up \ to\ O\left(\frac{1}{R^2}\right)
\cite{NS}.\ea By the same method used for the first tetrad given
by Eq. (37) we find that the momentum and angular momentum
associated with the second tetrad field given by Eq. (69) are \ba
\Pi^{(1) 1}(R,\theta,\phi)\A =\A 0, \qquad P^{(1)} =\int_VdV
\partial_{1}(\Pi^{(1)1}(R,\theta,\phi))=
\int_S dS_1 \Pi^{(1)1}(R,\theta,\phi) = 0, \nonu
  \Pi^{(2) 1}(R,\theta,\phi) \A=\A 0,
\quad P^{(2)}= 0 , \quad \Pi^{(3) 1}(R,\theta,\phi) = 0, \quad
P^{(3)}= 0 . \ea The non vanishing components of the
angular-momentum are given by \ba M^{(0)(1)}(e)\A=\A \frac{R
\sin\theta (\lambda(R)-4M^2e^{2\xi_0})}{4\pi \lambda(R)}\cong
 \frac{\sin\theta(R-M)}{4\pi}+O\left(\frac{1}{R}\right), \nonu
M^{(0)(1)}(E) \A \cong \A \frac{R
\sin\theta}{4\pi}+O\left(\frac{1}{R}\right), \nonu
M^{(0)(2)}(R,\theta,\phi) \A=\A \frac{MR^2 \cos\theta
\sqrt{(\lambda(R)-4M^2e^{2\xi_0})}}{4\pi \sigma(R)
\sqrt{\lambda(R)} },\nonu
M^{(0)(3)}(R,\theta,\phi) \A=\A
M^{(1)(2)}(R,\theta,\phi)=M^{(1)(3)}(R,\theta,\phi)=M^{(2)(3)}(R,\theta,\phi)=0.\ea
Using Eq. (74) in (65) we get
 \be L^{(0)(1)} = {\int_0^\pi}{\int_0^{2\pi}}{\int_{0}^\infty}
 d\theta d\phi dR \left[M^{(0)(1)}(e)-M^{(0)(1)}(E) \right] =
 M {\int_{0}^\infty} dR,\ee which give an infinite result!
  By  the same  method  we  can  obtain \be
 L^{(0)(2)} =  L^{(0)(3)} =
 L^{(1)(2)}=L^{(1)(3)} = L^{(2)(3)}= 0.\ee It is of interest to note that
the non-vanishing of  $L^{(0)(1)}$ is due to the appearance of
terms like   $\sin\theta$ while, the vanishing of  $L^{(0)(2)}$ is
due to the appearance of terms like   $\cos\theta$.

We show by explicit calculation that the energy-momentum tensor
which is a coordinate independent does not give  a consistent
result of the angular momentum when applied to the tetrad field
given by Eq. (23) which does not satisfy the boundary condition
given by Eq. (26).

\newsection{Main results and Discussion}
The main results of this paper are the following:\vspace{0.4cm}\\
$\bullet$  Tow different tetrad fields are used. The space-time
associated with these
tetrad fields is given by Eq. (25). \vspace{0.3cm}\\
$\bullet$  The energy of these tetrad fields are calculated using
the gravitational energy-momentum tensor, which is a coordinate
independent \cite{Mj}. One of this tetrad field given by Eq. (22)
gives a satisfactory results for the energy after using the
coordinate transformation given by Eq. (36). The other tetrad
field that is given by Eq. (23) its associated energy depends on
the radial coordinate.
\vspace{0.3cm}\\
$\bullet$ Calculations of the torsion components associated with
the two tetrad fields  are given. From these calculations we show
that the torsion components of each tetrad field are different.
This may gave an indication why the energy of the two tetrad
fields
is different. \vspace{0.3cm}\\
$\bullet$ We use the regularized expression of the gravitational
energy-momentum tensor to calculate the energy associated with the
second tetrad field given by Eq. (23).
 \vspace{0.3cm}\\
$\bullet$ We have shown that the energy associated with the second
tetrad field did not give the consistent  result even after using
the regularized expression of the gravitational energy-momentum
tensor. Therefore, we use the coordinate transformation given by
Eq. (36). Applying this coordinate transformation to the tetrad
field (23) we have got a satisfactory value of energy coincides
with the value of energy of the first tetrad field. \vspace{0.3cm}\\
$\bullet$ Using the definition of the energy and the angular
momentum given by Eqs. (13) and (20)   we show by explicit
calculations that the angular momentum  depends on the choice of
the frame used.\vspace{0.3cm}\\
$\bullet$ The calculation of the irreducible mass is given within
the external horizons using the Hamiltonian formulation. From this
calculation we show that the external horizons of each model does
not play any role on the energy. \vspace{0.3cm}\\ $\bullet$ We
have shown by explicit calculations that the diagonal tetrad field
which  given by Eq. (23) suffers from some
problems:\vspace{0.2cm}\\i) It does not satisfy the condition
given by Eq. (26) which guarantees the flatness of spacetime,
consequently the components of the torsion tensor did not vanish
when the physical quantities are set equal zero.\vspace{0.2cm}\\
ii) The use of the energy-momentum tensor given by Eq. (13)  did
not give the consistent results! Therefore,  we have used the
regularized expression of the energy momentum tensor and got the
consistent result for the energy. Also we have shown that both
expressions given by Eqs. (20) and (65) gave an infinite results
when  we have calculated the angular momentum  \cite{MTW}!\vspace{0.3cm}\\
$\bullet$ The construction of the tetrad given by Eq. (23) is the
square root of the metric given by Eq. (25) meanwhile, the
construction of the tetrad given by Eq. (22) is not the square
root of Eq. (25). Possible interpretations of the results given by
the second tetrad (which is the square root of the metric) is that
{\it it may  not be  a physical one}. Same problem has been
appeared \cite{MVR} for the Kerr solution. We need more studies to
confirm this conclusion.

\vspace{2cm} \centerline{\large {\bf Acknowledgment}} The author
would like to  thank the Referee for careful reading,  careful
checking the mathematics, putting the paper in  a more readable
form and the comments given for the second tetrad.

\end{document}